\documentclass[showpacs,prl,onecolumn,aps,superscriptaddress,preprintnumbers,letterpaper,11pt]{revtex4}
\usepackage{amsmath,amssymb}
\usepackage{epsfig}
\usepackage{graphicx}
\usepackage{amsmath}
\usepackage{amsfonts}
\usepackage{epstopdf}
\def\slashchar#1{\setbox0=\hbox{$#1$}     		
   \dimen0=\wd0                                 	
   \setbox1=\hbox{/} \dimen1=\wd1               	
   \ifdim\dimen0>\dimen1                        	
      \rlap{\hbox to \dimen0{\hfil/\hfil}}      	
      #1                                        	
   \else                                        	
      \rlap{\hbox to \dimen1{\hfil$#1$\hfil}}   	
      /                                         	
   \fi}

\newcommand{\be}{\begin{equation}}
\newcommand{\ee}{\end{equation}}
\newcommand{\bear}{\begin{eqnarray}}
\newcommand{\eear}{\end{eqnarray}}
\newcommand{\ba}{\begin{array}}
\newcommand{\ea}{\end{array}}

\begin{document}

\title{Anomalous soft photon production from the induced currents in Dirac sea}

\author{Dmitri E. Kharzeev}
\affiliation{Department of Physics and Astronomy, Stony Brook University, Stony Brook, New York 11794-3800, USA}
\affiliation{Department of Physics,
Brookhaven National Laboratory, Upton, New York 11973-5000, USA}

\author{Frash\"er Loshaj}
\affiliation{Department of Physics and Astronomy, Stony Brook University, Stony Brook, New York 11794-3800, USA}

\date{\today}

\begin{abstract}
The propagation of a high energy quark disturbs the confining QCD vacuum inducing the currents in Dirac sea.
Since quarks possess electric charge, these 
induced vacuum quark currents act as a source of soft photon radiation. This can lead to the enhancement of the soft photon production above the expectations based 
on the charged hadron yields and the Low theorem. 
We illustrate the phenomenon by using the exactly soluble $1+1$ dimensional 
massless Abelian gauge model that shares with QCD all of the ingredients involved in this mechanism: confinement, chiral symmetry breaking, axial anomaly, and the periodic $\theta$-vacuum. 
We show that the propagating quark throughout the process of hadronization induces in the vacuum charged transition currents that lead to a strong resonant enhancement of the soft photon yield; the Low theorem however remains accurate in the limit of very soft momenta.
We then construct on the basis of our result a simple phenomenological model and apply it to the soft photon production in the fragmentation of jets produced in $Z^0$ decays.  We 
find a qualitative agreement with the recent result from the DELPHI Collaboration.
 \end{abstract}

\maketitle

\section{\label{sec:intro}Introduction}
The production of soft photons in hadron collisions is governed by the Low theorem \cite{Low:1958sn} that is based on very general properties of QED as a vector gauge theory. In hadron collisions, the Low theorem states that soft photons are produced by the bremsstrahlung off the charged hadrons \cite{Gribov:1966hs}, and 
relates the soft photon yield to the measured hadron spectrum. This allows to make predictions that can be tested experimentally. Surprisingly, nearly every experiment that studied the production of soft photons in high energy hadron collisions found a dramatic (by factor of $2 \div 5$) enhancement above the Low theorem's predictions, see e.g. \cite{Chliapnikov:1984ed,Botterweck:1991wf,Banerjee:1992ut,Belogianni:1997rh,Belogianni:2002ib,Belogianni:2002ic} -- this is the long-standing puzzle of the ``anomalous photon production". Many theoretical models have been proposed to explain the anomalous photon production (for a review, see e.g. \cite{Balek:1989rx}); some of them are based on collective effects in produced hadronic matter \cite{Shuryak:1989vn,Lichard:1990ye}, including the effects of anomalies \cite{Basar:2012bp,Andrianov:2012hq,Fukushima:2012fg,Tuchin:2012mf,Yee:2013qma}; other invoke the synchrotoron radiation in the QCD vacuum \cite{Botz:1994bg}, new light bound states \cite{Wong:2010gf} or strong coupling phenomena treated within holography \cite{Hatta:2010kt} -- however none of them explains all features of the observed phenomenon.  
\vskip0.3cm
A particularly striking recent result is the measurement of the direct soft photon yield in  hadronic decays of the $Z^0$ boson by the DELPHI Collaboration  \cite{Abdallah:2010tk,Abdallah:2005wn}. 
The data recorded during the operation of the LEP $e^+ e^-$ collider at CERN show the photon spectrum similar to the one expected from bremsstrahlung, but with a magnitude about {\it four times} higher than the prediction of the Low theorem based on the measured charged hadron yields. Contrary to processes with hadronic final states, the DELPHI measurement \cite{Abdallah:2007aa} of photons produced in the $e^+ e^-\rightarrow Z^0\rightarrow \mu^+\mu^-+n \gamma$ channel is in good agreement with the theoretical expectation based on muon inner bremsstrahlung -- so the puzzle of the anomalous soft photons seems to be  specific to the production of hadronic final states. Remarkably, the soft photon yield was found \cite{Abdallah:2010tk} to be more sensitive to the neutral hadron multiplicity than to the charged one that is expected to govern the photon bremsstrahlung. 
\vskip0.3cm
The anomalous soft photons present a challenge to the foundations of  theory. Moreover, this puzzle is an obstacle to using soft photons for diagnosing quark-gluon plasma, as we need to understand the ``background" - the mechanism of photon production in elementary collisions. 
To be specific, in this paper we concentrate on the direct  soft photon production in hadronic decays of $Z^0$ that are dominated by 
the fragmentation of the quark jets produced in the $Z^0 \to q \bar{q}$ process. The direct (i.e. not originating from hadron decays) soft photon emission by the fragmenting quark and antiquark within the perturbative framework was found \cite{Abdallah:2010tk}  insufficient to explain the data. Since the bremsstrahlung of final state charged hadrons also does not describe the yield, one is naturally led to a possible nonperturbative mechanism of photon emission in the quark fragmentation process. 
Because of this, it is important to re-examine the problem by putting emphasis on the possible non-perturbative QCD effects on soft photon production. 
\vskip0.3cm
The quark propagating through the confining QCD vacuum pulls from the Dirac sea  the quark-antiquark pairs that later form hadrons; even if these hadrons are neutral, all quarks possess electric charge and can radiate photons. 
It is clear that such a mechanism involves QCD dynamics at large distances, where  the coupling is strong and 
we have to rely on an effective theory. Large energy of the fragmenting jets suggests using a dimensionally reduced 
theory describing the processes developing along the jet axis. Confinement is a crucial part of the mechanism that we study, and it has to be a part of the effective theory. 
 Based on the picture of confinement through the condensation of magnetic monopoles, which results in the quasi-Abelian projection \cite{'tHooft:1977hy,Mandelstam:1974pi}, we assume that the dynamics along the longitudinal direction is Abelian. Finally, to account for the hadronization process, we also need a confining theory in the presence of light fermions that severely affect the mechanism of confinement \cite{Gribov:1999ui}; for review, see \cite{Dokshitzer:2004ie}. Based on these assumptions, in \cite{Loshaj:2011jx,Kharzeev:2012re} we described the longitudinal dynamics of the jet by massless ${\rm QED}_2$, known as the Schwinger model \cite{Schwinger:1962tp,Lowenstein:1971fc,Coleman:1975pw}, which is exactly soluble. 
 Originally, the Schwinger model was used to discuss the qualitative features of jet fragmentation in \cite{:1974cks}. This model was also previously used to describe high energy processes in Refs. \cite{Fujita:1989vv,Wong:1991ub}. 
 An important property of this theory is that the particle production is determined by the axial anomaly as we will show below.

\section{\label{sec:topchocs}Oscillations of electric and axial charge in quark fragmentation}

\begin{figure}[htbp]
\centering
\includegraphics[width=.6\linewidth]{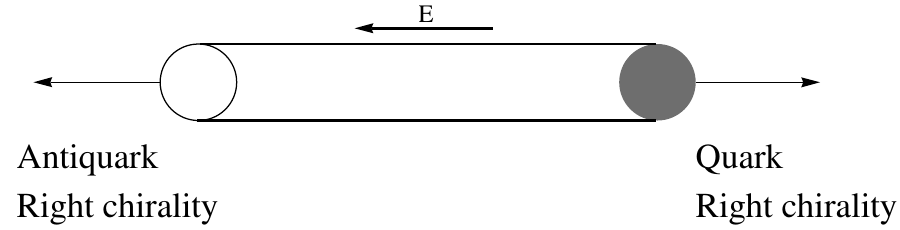} 
\caption{Fermion and antifermion moving back-to-back.}
\label{fig:btbchrg}
\end{figure}

The production of quark-antiquark pairs in the fragmentation of the quark allows interpretation in terms of topology 
of the gauge field and chirality of the quarks. In this section, we will illustrate this using the example of the $1+1$ dimensionally reduced theory.
As explained in the Introduction, we will assume that the dynamics of pair production along the jet axis can be modeled by massless Quantum Electrodynamics in $1+1$ dimensions or ${\rm QED}_2$:
\begin{equation}
\mathcal{L}=-\frac{1}{4}G_{\mu\nu} G^{\mu\nu}+\bar{\psi}i\gamma^\mu \partial_\mu \psi-g \bar{\psi}\gamma^\mu \psi B_\mu - g j_{ext}^\mu B_\mu
\label{eq:qed2}
\end{equation}
where $\mu=0,1$, $B_\mu$ is the gauge field, $G_{\mu\nu}=\partial_\mu B_\nu-\partial_\nu B_\mu$ is the field strength and $j_{ext}^\mu$ is an external current. As in \cite{Loshaj:2011jx,Kharzeev:2012re}, the leading quarks of the jets are introduced through an external current composed by the fermion and the anti-fermion moving back-to-back with equal velocities $v$ (see Fig.\ref{fig:btbchrg}):

\begin{equation}
j_{ext}^0(x)=\delta(z-vt)\theta(z)-\delta(z+vt)\theta(-z)
\label{eq:ec}
\end{equation}
where $x^\mu=(t,z)$ and 
\begin{equation}
v=\frac{p_{jet}}{\sqrt{p_{jet}^2+Q_0^2}}
\label{eq:vel}
\end{equation}
where $p_{jet}$ is the jet momentum and $Q_0 \sim 2$ GeV is the time-like virtuality scale at which the pQCD DGLAP cascade stops, and the effects of confinement described by our effective theory begin to operate. 
It is well known that the Schwinger model can be solved by bosonisation \cite{Coleman:1974bu,Mandelstam:1975hb}. 
For the vector current, the bosonization relation is
\begin{equation}
j^\mu(x) = \bar{\psi}(x)\gamma^\mu \psi(x)=-\frac{1}{\sqrt{\pi}}\epsilon^{\mu\nu}\partial_\nu\phi(x)
\label{eq:bc}
\end{equation}
where $\phi$ is a real scalar field; note that with (\ref{eq:bc}) the conservation of vector current is automatic.
It is easy to express the axial current in terms of the scalar field as well -- in $1+1$ dimensions, $\gamma^\mu \gamma^5=-\epsilon^{\mu\nu}\gamma_\nu$, therefore 
\begin{equation}
j_5^\mu(x)=\bar{\psi}(x)\gamma^\mu \gamma^5  \psi(x)=\frac{1}{\sqrt{\pi}}\partial^\mu \phi(x) .
\label{eq:acurr}
\end{equation}
Using the bosonization relations above, one can see that the original ${\rm QED}_2$  acquires the form of a theory describing a real massive scalar field coupled to a classical source $\phi_{ext}$ that is given by
\begin{equation}
j_{ext}^\mu(x)=-\frac{1}{\sqrt{\pi}}\epsilon^{\mu\nu}\partial_\nu\phi_{ext}(x) .
\label{eq:bec}
\end{equation}
The equation of motion for the scalar field is \cite{:1974cks,Loshaj:2011jx,Kharzeev:2012re}
\begin{equation}
(\Box+m^2)\phi(x)=-m^2\phi_{ext}(x)
\label{eq:eom}
\end{equation}
where $\Box \equiv \partial_t^2 - \partial_z^2$ and $m^2=g^2/\pi$; note that the coupling constant $g$ has dimension of mass in $1+1$ dimensions. 
\vskip0.3cm
The coupling to a classical source results in particle creation. The produced particles (that after bosonization become the quanta of the $\phi$ field) can be interpreted as neutral mesons produced in the fragmentation of the string stretched between the original 
quark and antiquark. 
Let us for a moment consider the ultra-relativistic limit $v\rightarrow 1$ in which the problem becomes somewhat simpler. Due to the Lorentz invariance, we can write the equation of motion in terms of the proper time $\tau$ and rapidity $y$: 
\begin{eqnarray}
\tau&=&\sqrt{t^2-z^2} \nonumber \\
y&=&\frac{1}{2}\ln\frac{t+z}{t-z} .
\label{eq:ty}
\end{eqnarray}
The equation of motion \eqref{eq:eom} in the $v\rightarrow 1$ limit is independent of rapidity and can be written as
\begin{equation}
(\Box+m^2)\phi(m \tau)=-m^2\phi_{ext}(m \tau) ,
\label{eq:eom1}
\end{equation}
which is an ordinary differential equation in the proper time:
\begin{equation}
(\Box+m^2)\phi(m \tau)=m \frac{1}{\tau}\phi'(m \tau)+m^2 \phi''(m \tau)+m^2\phi(m \tau)=-m^2\phi_{ext}(m \tau) ,
\label{eq:eom2}
\end{equation}
where the prime denotes differentiation with respect to $m\tau$.
On the other hand, 
\begin{equation}
\phi_{ext}(m \tau)=-\sqrt{\pi}\int^z{dz'}[-\delta(z+t)\theta(-z)+\delta(z-t)\theta(z)]=\sqrt{\pi}\theta(t^2-z^2)=\sqrt{\pi}\theta(\tau^2) ;
\label{eq:}
\end{equation}
we therefore have to solve
\begin{equation}
\phi''(m \tau)+\frac{1}{m \tau}\phi'(m \tau)+\phi(m \tau)=-\sqrt{\pi}\theta(m^2 \tau^2) .
\label{eq:eom3}
\end{equation}
 The solution can be written as
\begin{equation}
\phi(m \tau)=-\sqrt{\pi}\theta(m^2 \tau^2)(1-J_0(m \tau)),
\label{eq:eomsol}
\end{equation}
where $J_0$ is the Bessel function. 
The equations \eqref{eq:eomsol} and \eqref{eq:bc} show that the evolution in proper time gives rise to oscillation in the vector (electric) charge density
\begin{eqnarray}
j^0(\tau,y)&=&\frac{1}{\sqrt{\pi}}\partial_z(\sqrt{\pi}\theta(m^2 \tau^2)(1-J_0(m \tau)))=-\partial_z J_0(m\tau) \nonumber \\
&=&-(-m\sinh y (-J_1(m\tau)))=-m\sinh y J_1(m\tau) .
\label{eq:}
\end{eqnarray} 
It is this oscillation of electric charge that will be responsible for the enhancement of the soft photon yield once we introduce the coupling to the dynamical $(3+1)$ dimensional electromagnetic field.
\vskip0.3cm
Since the oscillation of electric charge is crucial for our interpretation of the anomalous soft photon production, it is worthwhile to discuss this phenomenon in more detail. We will now show that the oscillation of electric charge is induced by the axial anomaly in the presence of chirality imbalance \footnote{In $(3+1)$ dimensions, these are the crucial ingredients of the Chiral Magnetic Effect \cite{CME}.}.
In $(1+1)$ dimensions, the helicity of a fermion is determined simply by the direction of its motion -- the fermion moving to the right is right-handed, and the fermion moving to the left is left-handed. 
For an antifermion, just like in $(3+1)$ dimensions, chirality and helicity have the opposite signs - so the antifermion moving to the left has right-handed chirality. 
Our original configuration of a fermion-antifermion pair moving back-to-back therefore has two units of chirality, see Fig.1.
\vskip0.3cm
The index theorem in two dimensions is given by
\begin{equation}
\Delta Q_5=N_R-N_L=\frac{g}{\pi}\int{d^2 x}F_{01}
\label{eq:it}
\end{equation}
where $N_{R,L}$ is the number of left/right movers. We can see that a nonzero axial charge gives rise to an electric field. This has a simple physical interpretation -- the presence of electric field due to the Lorentz force creates an asymmetry between the left- and right-moving charged fermions.
Using bosonization relations and requiring that the fields vanish at infinity, we can find the electric field induced by the pairs created by the original external source: 
\begin{equation}
F^{ind}_{01}=-\frac{g}{\sqrt{\pi}}\phi .
\label{eq:ind}
\end{equation}
\vskip0.3cm
Let us now show that the axial charge $\Delta Q_5$ oscillates as a function of time. To do this, we will use \eqref{eq:it}, where $F_{01}^{tot} = F^{ind}_{01} + F_{01}^{ext}$ will be the sum of the electric field (\ref{eq:ind}) induced by the pair creation and the field created by the external source:
\begin{equation}
F^{ext}_{01}=-g[\theta(z+t)-\theta(z-t)]=-g\theta(t^2-z^2) .
\label{eq:ees}
\end{equation}
The total axial charge as a function of time is thus given by
\begin{eqnarray}
\Delta Q_5&=&\frac{g}{\pi}\int{d^2 x}F_{01}^{tot}=\frac{g}{\pi}\int{d^2 x}\left(-g\theta(m^2 \tau^2)+g\theta(m^2 \tau^2)(1-J_0(m \tau))\right)\nonumber \\
&=&-\frac{g^2}{\pi}\int_0^t{dt'}\int_{-t'}^{t'}{dz}J_0(m \sqrt{t'^2-z^2})=-2m^2\int_0^t{dt'}\int_{0}^{t'}{dz}J_0(m \sqrt{t'^2-z^2}) \nonumber \\
&=&2m\int_0^t{dt'}\sin(mt')=2[\cos(mt)-1]
\label{eq:}
\end{eqnarray}
The axial charge thus indeed oscillates with the period $T=2\pi/m$ -- the appearance of $m$ of course is not surprising since it is the only scale in the theory. However the oscillation of the axial charge is a non-trivial consequence of i) the periodicity of the $\theta$-vacuum of the Schwinger model, and ii) non-equilibrium nature of our process. The chiral charge of the original quark-antiquark pair with time is screened by the electric field, which then consequently decays producing additional chiral quark-antiquark pairs -- because the set-up describing the separating quark and antiquark jets is far from equlibrium, the axial charge keeps oscillating around its equilibrium value of $\Delta Q_5 = 0$. Because the axial anomaly in $(1+1)$ dimensions couples the axial and vector currents, as indicated by (\ref{eq:bc}) and (\ref{eq:acurr}), the oscillations of axial charge translate into the oscillations of electric current\footnote{This is the $(1+1)$ analog of the Chiral Magnetic Wave \cite{Kharzeev:2010gd} -- a gapless collective mode induced by the axial anomaly in $(3+1)$ dimensional hydrodynamics.}. We will now couple our theory to the $(3+1)$ electromagnetic field, and show 
that the fluctuations of electric current induced by the anomaly indeed source the soft photon production. 
\section{\label{sec:phot}The soft photon production due to the axial anomaly}
In the previous section we have considered the sources moving along the lightcone, and have demonstrated that the axial anomaly leads to the undamped axial charge oscillations with frequency $m$ (the mass of the scalar meson). 
Let us now consider a more realistic case when the sources move with velocity $v<1$ given by \eqref{eq:vel}. We will  solve equation $\eqref{eq:eom}$ and using the bosonization relation \eqref{eq:bc} get the total electric current induced by the quark-antiquark pairs. We then couple this current to $(3+1)$ electromagnetic field and compute the rate of the soft photon bremsstrahlung. 
\vskip0.3cm
The general solution to the equation of motion \eqref{eq:eom} is given by
\begin{equation}
\phi(x)=\phi_0(x)+i\int{d^2x'}D_R(x-x')(-m^2\phi_{ext}(x')) ,
\label{eq:eoms}
\end{equation}
where $D_R(x)$ is the retarded scalar field propagator and $\phi_0(x)$ satisfies the Klein-Gordon equation. 
Taking the Fourier transform of \eqref{eq:bec}
\begin{equation}
\tilde{j}^\mu_{ext}(p)=-\frac{1}{\sqrt{\pi}}\epsilon^{\mu\nu}(-ip_\nu)\tilde{\phi}_{ext}(p) ,
\label{eq:fbec}
\end{equation}
we can solve for $\tilde{\phi}_{ext}(p)$
\begin{equation}
\tilde{j}^0_{ext}(p)=-\frac{1}{\sqrt{\pi}}(-ip_1)\tilde{\phi}_{ext}(p)\Rightarrow \tilde{\phi}_{ext}(p)=-i\frac{\sqrt{\pi}}{p_1}\tilde{j}^0_{ext}(p) .
\label{eq:pt}
\end{equation}
Let us recall that the direction of $p^1$ is along the jet axis, which we choose to be the $z$ direction, therefore $p^1=-p_1 \equiv p_z$. We are only interested in the contribution to \eqref{eq:eoms} which is induced by the interaction with the external source; therefore we can write the solution in momentum space as
\begin{equation}
\tilde{\phi}(p)=\frac{-m^2\tilde{\phi}_{ext}(p)}{-p_\mu p^\mu+m^2}=-m^2\left(i\frac{\sqrt{\pi}}{p_z}\tilde{j}^0_{ext}(p)\right)\frac{1}{-p_\mu p^\mu+m^2} ;
\label{eq:kgep}
\end{equation}
this is the scalar field induced by the external source. We will need the Fourier transform of \eqref{eq:ec} and it is given by
\begin{equation}
\tilde{j}_{ext}^0(p)=i\frac{2vp_z}{p_0^2-v^2 p_z^2} .
\label{eq:j0}
\end{equation}
We also need $j^1_{ext}$, which can be computed using the conservation equation $\partial_0 j_{ext}^0=-\partial_1 j_{ext}^1$ :
\begin{equation}
\tilde{j}_{ext}^1(p)=-\frac{p^0}{p_1}\tilde{j}_{ext}^0(p)=i\frac{2 v p^0}{p_0^2-v^2 p_z^2} .
\label{eq:j1}
\end{equation}
The induced meson field in momentum space can now be writen down as
\begin{equation}
\tilde{\phi}(p)=\frac{-1}{p_\mu p^\mu -m^2}\frac{2\sqrt{\pi}m^2 v}{p_0^2-v^2 p_z^2} .
\label{eq:pf}
\end{equation}
By taking the inverse Fourier transform of \eqref{eq:pf} and using the bosonization relation $\eqref{eq:bc}$, we can compute the induced vector current $j^\mu(x)$. 
\vskip0.3cm
So far we have been considering only the strong interaction dynamics that within a jet was modeled by a $(1+1)$ dimensionally reduced effective theory. Since we know that all quarks possess in addition to color also the electric charge, they couple to electromagnetic fields, with a different coupling constant $\sim e$. The electromagnetic field of course is not confined within a string (unlike the gauge field $B_\mu$ that we considered so far) and so can propagate in $(3+1)$  dimensions. Therefore we have to couple the total current $j_{tot}$ of the $(1+1)$ quarks to the $(3+1)$ dimensional electromagnetic gauge field $A_\mu$. Our system thus resembles a quantum wire where the charges propagate only along the wire but can radiate photons in $(3+1)$ dimensions. The resulting theory has a familiar form
\begin{equation}
\mathcal{L}=-\frac{1}{4}F_{\mu\nu}F^{\mu\nu} +  j_{tot}^\mu A_\mu .
\label{eq:qed}
\end{equation}
The photon bremsstrahlung spectrum can now be evaluated using the standard formula
\begin{equation}
\frac{dN_\gamma}{d^3 p}=\frac{1}{(2\pi)^3}\frac{1}{2 p^0}|\tilde{j}_{tot}^\mu(p) \tilde{j}^*_{tot,\mu}(p)| ,
\label{eq:tpp}
\end{equation}
where $\tilde{j}_{tot}$ is the Fourier transform of the current. In $3+1$ dimensions, the photon distribution will depend on the $(3+1)$ Fourier transform of the current, therefore we make the identification $p_\mu p^\mu=p_0^2-p^2_z= p_\perp^2$. 
The contribution from the sea quarks (i.e. the quarks produced from the vacuum by the external source) is given by
\begin{equation}
\tilde{j}^\mu_{\mathrm{sea}}(p)=-eQ_f \frac{1}{\sqrt{\pi}}\epsilon^{\mu\nu}(-i p_\nu)\tilde{\phi}(p)=-i eQ_f \frac{\epsilon^{\mu\nu}p_\nu}{p_\perp^2-m^2}\frac{2m^2 v}{p_0^2-v^2 p_z^2}
\label{eq:j22}
\end{equation}
where $Q_f$ is the fraction of the electric charge for a given quark flavor. Note that the spatial component is now in the $z$ direction, so in \eqref{eq:j22} we should take $\epsilon^{03}=-\epsilon^{30}$=1. We should also add to \eqref{eq:j22} the contribution from valence quarks -- the original quark and antiquark moving back-to-back along the $z$ direction with velocities $v$. Their current is given by
\begin{eqnarray}
j_{\mathrm{val}}^0(x)&=&eQ_f\delta(x)\delta(y)[\delta(z-vt)\theta(z)-\delta(z+vt)\theta(-z)] \nonumber \\
j_{\mathrm{val}}^3(x)&=&eQ_f\delta(x)\delta(y)[v\delta(z-vt)\theta(z)+v\delta(z+vt)\theta(-z)]
\label{eq:j03}
\end{eqnarray}
We now take the Fourier transform
\begin{equation}
\tilde{j}^0_{\mathrm{val}}(p) = e \int{d^4 x} e^{i p\cdot x} j_{\mathrm{val}}^0(x)=^{\epsilon\rightarrow 0}e\int_0^\infty{dt}e^{i p^0 t-\epsilon t}\left(e^{-ip_z v t}-e^{ip_z v t}\right)=i eQ_f\frac{2 v p_z}{p_0^2-v^2 p_z^2} .
\label{eq:}
\end{equation}
Similarly, 
\begin{equation}
\tilde{j}_{\mathrm{val}}^3(p)=ie\frac{2 v p^0}{p_0^2-v^2 p_z^2} .
\label{eq:}
\end{equation}
The total current that contributes to the photon production is given by the sum of the sea and valence contributions: 
\begin{equation}
\tilde{j}_{\mathrm{tot}}^\mu(p)=\tilde{j}_{\mathrm{sea}}^\mu(p)+\tilde{j}_{\mathrm{val}}^\mu(p) .
\label{eq:totcurr}
\end{equation} 
From \eqref{eq:tpp} and \eqref{eq:totcurr}, we can now compute the photon spectrum. Let us consider the case of $Z_0$ decay to quarks; to do so we have to introduce the probability for $Z_0$ to decay to a certain flavor of quark. The final formula for the photon spectrum is thus given by
\begin{eqnarray}
\frac{dN_\gamma}{d^3 p}&=&\left(\frac{\Gamma_{uu}+\Gamma_{cc}}{\Gamma_{\mathrm{hadron}}}\left(\frac{2}{3}\right)^2+\frac{\Gamma_{dd}+\Gamma_{ss}+\Gamma_{bb}}{\Gamma_{\mathrm{hadron}}}\left(\frac{1}{3}\right)^2\right)\frac{1}{(2\pi)^3}\frac{1}{2 p^0}e^2 \frac{4 v^2}{(p_0^2-v^2 p_z^2)^2}p_\perp^2\left(1+\frac{m^2}{p_\perp^2-m^2}\right)^2 \nonumber \\
&=&\left(B_{2/3}\left(\frac{2}{3}\right)^2+B_{1/3}\left(\frac{1}{3}\right)^2\right)\frac{1}{(2\pi)^3}\frac{1}{2 p^0}e^2 \frac{4 v^2}{(p_0^2-v^2 p_z^2)^2}p_\perp^2\left(1+\frac{m^2}{p_\perp^2-m^2}\right)^2
\label{eq:tpp1}
\end{eqnarray}
where $\Gamma_{ff}$ is the decay width of $Z_0$ to quark-antiquark of flavor $f$ and $\Gamma_{\mathrm{hadron}}$ is the total decay width of $Z_0$ to hadrons; the Particle Data Group \cite{Beringer:1900zz} gives the values $B_{2/3}=0.331$ and $B_{1/3}=0.669$. 
\vskip0.3cm
 The formula $\eqref{eq:tpp1}$ is the central result of our paper. Without the second term in the parenthesis, it is just the usual formula for the bremsstrahlung radiation off the original quark and antiquark produced by the $Z_0$ decay. The second term in the parenthesis is the contribution of the quantum back-reaction of the vacuum  -- in other words, this term represents the photons produced by the transient quark-antiquarks pairs created in the fragmentation of the string. This term originates from  the scalar propagator of the field $\phi$ which in our model is stable, so we get a spectral density of an infinitely narrow resonance -- so there is a sharp resonance in photon production at $p_\perp=m$, the frequency of the vacuum current oscillation. It is important to note that in the soft photon limit of $p_\perp \to 0$, the two terms in the parenthesis cancel each other -- this is in accord with the Low theorem stating that the very soft photons can be produced only by the asymptotic states -- the charged mesons. In our case, the transient quarks and antiquarks are ultimately bound into neutral mesons and so are not allowed to contribute to the photon spectrum in the $p_\perp \to 0$ limit.
 
\section{Phenomenology of soft photon production}

\begin{figure}
\centering
\includegraphics[width=0.7\linewidth]{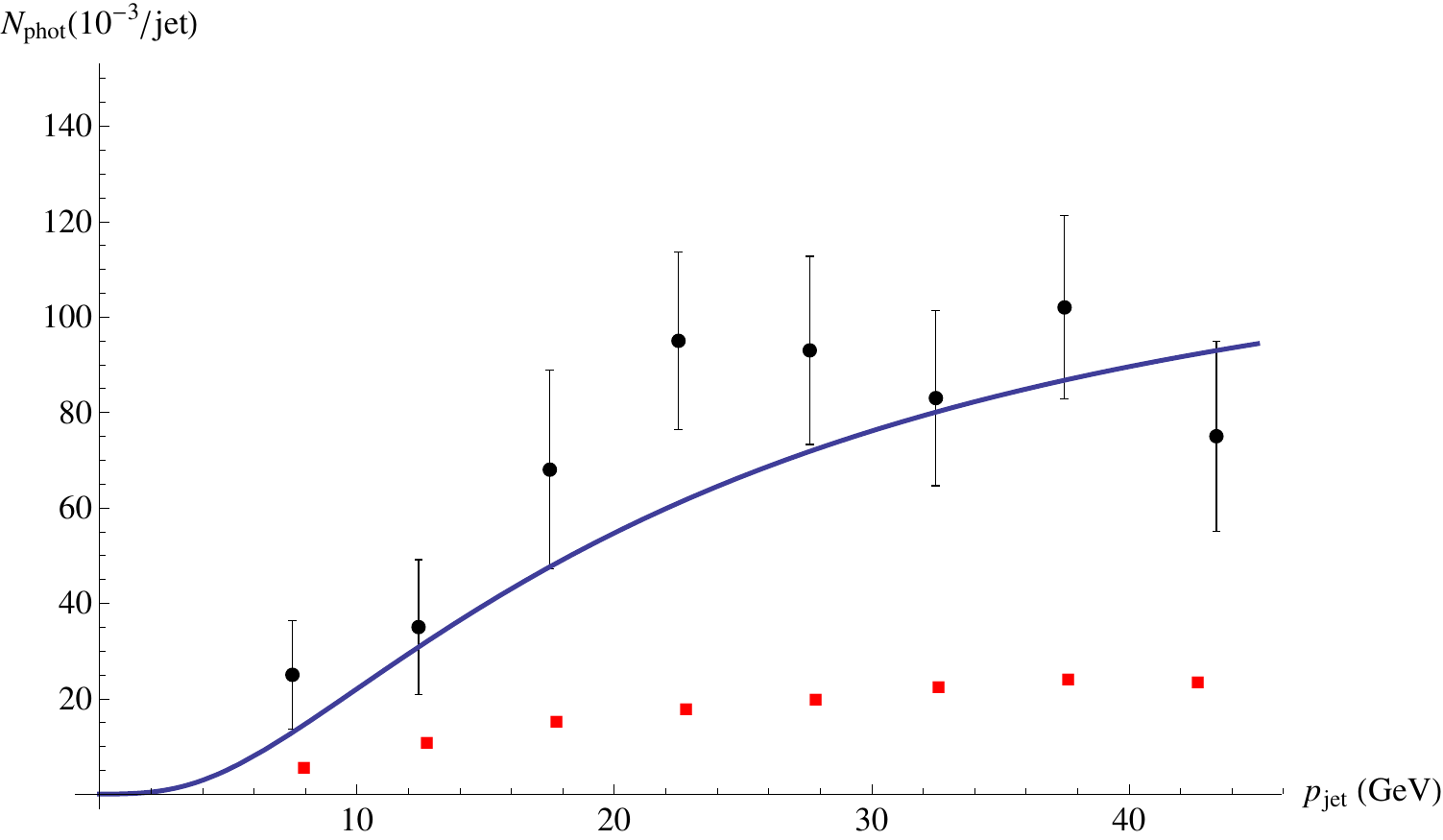} 
\caption{The soft photon yield as a function of the jet momentum. The circles are the DELPHI Collaboration data \cite{Abdallah:2010tk}, and the squares represent the conventional theoretical calculations of the soft photon bremsstrahlung \cite{Abdallah:2010tk}. The solid line is our result.}
\label{fig:ngvspjet}
\end{figure}

It is clear that our model is unrealistic in assuming the zero width of the meson $\phi$ -- all mesons that exist in the hadron spectrum possess non-zero width. Moreover, the mass $m$ in reality cannot be a fixed number, as the scalar meson of our effective theory represents the entire hadron spectrum. 
To make our model more realistic we thus have to i) consider a distribution in $m$ and ii) to account for a finite width of the mesons. 
To reach the objective i), let us first consider the potential acting between the static fermions in the Schwinger model that we use. For a fermion-antifermion pair separated by the distance $r$, the potential in this model is given by \cite{Lowenstein:1971fc}
\begin{equation}
V(r)=\frac{g \sqrt{\pi}}{2}\left(1-e^{-\frac{g}{\sqrt{\pi}}r}\right) .
\end{equation}
At large distances, the potential is screened by the produced pairs, but at short distances $r \ll m^{-1}$ the potential is linear, 
\begin{equation}
V(r) \simeq \frac{g\sqrt{\pi}}{2}\frac{g}{\sqrt{\pi}}r
=\frac{g^2}{2}r=\frac{\pi}{2}m^2 r, \ \ r \ll m^{-1} ,
\label{eq:pot}
\end{equation}
with the string tension $\kappa^2=\frac{\pi}{2}m^2$.  To introduce a distribution in $m$, let us 
assume that the string tension fluctuates with a Gaussian probability distribution \cite{Bialas:1999zg}
\begin{equation}
P(\kappa^2)=\sqrt{\frac{2}{\pi <\kappa^2>}}e^{-\frac{\kappa^2}{2 <\kappa^2>}} ,
\label{eq:}
\end{equation} 
where 
\begin{equation}
<\kappa^2>=\int_0^\infty{d\kappa P(\kappa^2)\kappa^2} .
\label{eq:}
\end{equation}
We use the mean value of the string tension $<\kappa^2> = 0.9$ GeV/fm suggested by the lattice studies and Regge phenomenology.
\vskip0.3cm
To account for a finite decay width of the meson, we write the propagator as
\begin{equation}
\frac{1}{p_\perp^2-m^2}\rightarrow \frac{1}{p_\perp^2-m^2+i\gamma^2}
\label{eq:prop1}
\end{equation}
where $\gamma$ is an effective width. From (\ref{eq:tpp1}) it is clear that the soft photon production will be dominated by the longest living resonances.
Using the PDG \cite{Beringer:1900zz}  values for the masses and widths of the neutral isoscalar resonances, we find the values of $\gamma = \sqrt{m \Gamma}$ in the range of $\gamma \simeq 8\ 10^{-4}$ GeV for the $\eta$ meson, and $\gamma \simeq 8\ 10^{-2}$ GeV for the $\omega$ meson. We will see 
that the value of $\gamma$ extracted from the fit to the DELPHI data on soft photon production falls in this range.
\vskip0.3cm
The DELPHI Collaboration \cite{Abdallah:2010tk} measured the photons with transverse momenta $p_\perp<80$ MeV and total energies within $0.2<E_\gamma<1$ GeV. 
We thus substitute \eqref{eq:prop1} in \eqref{eq:tpp1} and compute the total number of photons in this kinematic domain 
\begin{equation}
N_\gamma =\int{dm \ \sqrt{\frac{\pi}{2}} \ P(\frac{\pi}{2}m^2) \ \left(\int{d^3 p \frac{dN}{d^3p}}\right) }  
\label{eq:total}
\end{equation} 
by integrating over the appropriate range of transverse momentum and energy. 
Since $\gamma$ is small, we can use  
\begin{equation}
\delta(x)=\lim_{\epsilon\rightarrow 0}\frac{1}{\pi}\frac{\epsilon}{x^2+\epsilon^2}
\label{eq:}
\end{equation}
and write 
\begin{equation}
\left|1+\frac{1}{p_\perp^2-m^2+i\gamma^2}\right|^2=\frac{p_\perp^4+\gamma^4}{(p_\perp^2-m^2)^2+\gamma^4}\rightarrow \left(1+\frac{p_\perp^4}{\gamma^4}\right)\left(\gamma^2\frac{ \pi}{2m}\right)\delta(p_\perp-m)
\label{eq:df}
\end{equation}
We use the delta function to eliminate the integral over $p_\perp$ in (\ref{eq:total}). We use the standard value of the string tension $<\kappa^2>=0.9$ GeV/fm and extract the value of the parameter $\gamma = 0.003$ GeV by fitting to the measured experimental photon yield. 
In Fig. \ref{fig:ngvspjet} we show the result as compared to the DELPHI data on the total number of photons as a function of the jet momentum \cite{Abdallah:2010tk}. One can see that our mechanism describes the observed enhancement reasonably well, with the fitted value of the parameter $\gamma = 0.003$ GeV within a reasonable range expected for neutral isoscalar resonances.  
\section{\label{sec:concl}Summary and outlook}

We have modeled the propagation of a high energy quark through the confining QCD vacuum by using an exactly soluble $1+1$ dimensional 
massless Abelian gauge model. While this is of course a drastic oversimplification of the problem, it is important to keep in mind that this model
shares with QCD all of the ingredients involved in the mechanism that we propose: confinement, chiral symmetry breaking, axial anomaly, and the periodic $\theta$-vacuum. 
Moreover, at high energies the factorization of the transverse and longitudinal degrees of freedom is natural, and so the use of a dimensionally reduced theory may be justified. In this theory, we have established the phenomenon of coherent oscillations of the axial and vector (electric) charges coupled by the axial anomaly and induced by the propagating high energy quark. These oscillations originate from the continuous production of quark-antiquark pairs pulled from the vacuum to screen the axial and electric charges of the external source. We have found that soft photons provide an important signature of this mechanism, as it leads to a strong enhancement of the soft photon yield. At the cost of introducing an adjustable parameter, our model can then describe the DELPHI data on the soft photon production. 
\vskip0.3cm
We readily admit that the use of a $1+1$ model, and the procedure we use to compare our results to the experimental data can be questioned. Moreover, 
our numerical result depends on an adjustable parameter that cannot be determined within our model (even though its value appears reasonable). 
Nevertheless, we hope that the described mechanism of string fragmentation may be close to the one in real $(3+1)$ QCD. It would be interesting to 
generalize our study to a $(3+1)$ model as it would allow, for example, to address the effects of spin and chirality on the fragmentation of a polarized quark (see \cite{Kang:2010qx} for an attempt in this direction). 

\vskip0.2cm
{\bf Acknowledgements.}
We thank Yuri Dokshitzer for stimulating this study by bringing the soft photon puzzle in $Z$ decays to our attention. We are grateful to 
Joshua Ilany for participating in the early stage of this work, and to Edward Shuryak, Alexander Stoffers, and Ismail Zahed for useful discussions. 
This work was supported in part by the U.S. Department of Energy under Contracts No.
DE-FG-88ER40388 and DE-AC02-98CH10886.

\bibliographystyle{unsrt}

\end{document}